\documentclass[aps,pra,twocolumn,groupedaddress,floatfix]{revtex4-1}
\usepackage{textcomp}
\usepackage{amsmath}
\usepackage{graphicx}
\usepackage{color}
\usepackage[colorlinks=true,citecolor=blue,urlcolor=blue,linkcolor=blue]{hyperref}
\usepackage[all]{hypcap}

\begin{document}


\title{Paschen-Back effect and Rydberg-state diamagnetism in vapor-cell  electromagnetically induced transparency}


\author{L.~Ma$^1$}
\author{D.~A.~Anderson$^2$}
\author{G.~Raithel$^{1,2}$}
\affiliation{1. Department of Physics, University of Michigan, Ann Arbor, Michigan 48109}
\affiliation{2. Rydberg Technologies LLC, Ann Arbor, Michigan 48104}



\date{\today}

\begin{abstract}
We report on rubidium vapor-cell Rydberg electromagnetically induced transparency (EIT) in a 0.7~T magnetic field where all involved levels are in the hyperfine Paschen-Back regime, and the Rydberg state exhibits a strong diamagnetic interaction with the magnetic field. Signals from both $^{85}\mathrm{Rb}$ and $^{87}\mathrm{Rb}$ are present in the EIT spectra. This feature of isotope-mixed Rb cells allows us to measure the field strength to within a $\pm 0.12$\% relative uncertainty. The measured spectra are in excellent agreement with the results of a Monte Carlo calculation and indicate unexpectedly large Rydberg-level dephasing rates. Line shifts and broadenings due to small inhomogeneities of the magnetic field are included in the model.
\end{abstract}


\maketitle


\section{INTRODUCTION}
Electromagnetically induced transparency (EIT) is a quantum interference
process where two excitation pathways in a three-level atomic structure
destructively interfere and produce an increase in the transmission of one
of the utilized laser beams~\cite{PhysRevLett.66.2593,RevModPhys.77.633}. In the Rydberg-EIT cascade scheme~\cite{PhysRevLett.98.113003}, the transparency
is formed by a coherent superposition of the ground state and the
Rydberg state. Rydberg-EIT has been implemented in both cold atomic
gases~\cite{Gavryusev2016,Weatherill2008} and in room-temperature
vapor cells~\cite{PhysRevLett.98.113003,PhysRevA.87.042522}. It has been
widely used as a nondestructive optical detection
technique for Rydberg spectroscopy~\cite{PhysRevA.83.052515,Grimmel2015},
quantum information processing~\cite{RevModPhys.82.2313} and measurements
of weak~\cite{Sedlacek2012819,Holloway20146169} and strong~\cite{PhysRevApplied.5.034003,Miller2016} microwave electric fields. Recently, EIT in vapor cells has been employed to investigate Cs Rydberg atoms in magnetic fields up to $\sim$~$100~\mathrm{G}~$\cite{PhysRevA.94.043822} and Rb $5D_{5/2}$ atoms in fields up to $\sim$~$0.6~\mathrm{T}~$\cite{PhysRevA.93.043854,Whiting:arXiv2016}. 

In this work, we investigate Rydberg atoms in a 0.7~T magnetic field. In magnetic fields $B$ exceeding $2n^4$ at.un. (for $n=33, B=0.4~\mathrm{T}$), where $n$ is the principal quantum number, the diamagnetic term dominates and mixes states with different angular momentum quantum numbers~\cite{gallagher:2005rydberg}. We employ the $\lvert33S_{1/2},m_{j}=1/2\rangle$ Rydberg state whose interaction with the B field includes a Zeeman term (linear) and
a  diamagnetic term (quadratic). In a 0.7~Tesla field, the diamagnetic interaction accounts for about 70\% of the differential magnetic dipole moment of $3.06~\mu_B$. In this field, both the ground and intermediate states are in the hyperfine Paschen-Back regime, and the energy separations between the magnetic states are larger than the spectroscopic Doppler width.

We demonstrate that the presented method offers two advantages in high-magnetic-field measurements. First, the diamagnetic interaction gives rise to an enhanced differential dipole moment, enabling measurement of small magnetic field changes on a large magnetic field background. Second, simultaneous measurements of  field-induced level shifts for both $^{85}\mathrm{Rb}$ and $^{87}\mathrm{Rb}$ isotopes affords high absolute accuracy in magnetic field measurements based on relative line separations.

\section{EXPERIMENTAL SETUP}
\label{sec:2}
\begin{figure}
\begin{centering}
\includegraphics[width=\columnwidth]{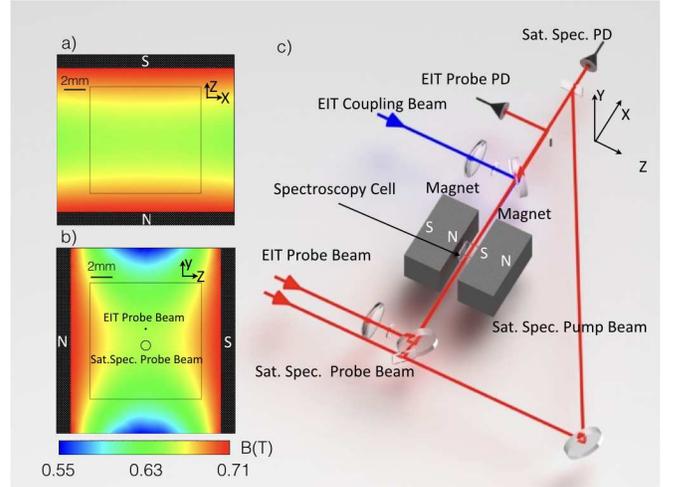}
\par\end{centering}
\caption{(color online) Calculation of the magnetic field $B$ in
the $x$-$z$ plane (a) and $y$-$z$ plane (b). The filled black regions represent the
poles of the bar magnets. The open black square represents the inner boundary
of the spectroscopy cell. (c) Illustration of the experimental setup, including an EIT channel and a saturation spectroscopy channel. The two channels are separated in the $y$ direction.
\label{fig:1}}
\end{figure}

The experimental setup is illustrated in Fig.~\ref{fig:1}.
The magnetic field in our experiment is produced by two N52 Neodymium
permanent magnets. The
$B$ field is calculated using a finite-element
analysis software (ANSYS Maxwell). Figures~\hyperref[fig:1]{\ref*{fig:1}(a)}
and \hyperref[fig:1]{\ref*{fig:1}(b)} show two cuts of the magnetic field.  A spectroscopic cell filled with a natural Rb isotope mixture is placed between the magnets. In order to increase the optical absorption in the cell, the cell temperature is maintained at $\sim45$~\textcelsius{} during the measurements by heating both the cell and surrounding magnets. 

The optical setup includes two measurement channels: a Rydberg-EIT channel and a saturation spectroscopy (Sat. Spec.) channel. As shown in Fig.~\hyperref[fig:1]{\ref*{fig:1}(b)} and~\hyperref[fig:1]{\ref*{fig:1}(c)}, the two channels are parallel to the $x$-axis and separated by $1.85~\mathrm{mm}$ in the $y$-direction. The Rydberg-EIT probe beam is focused to a waist of $\sim40~\mu\mathrm{m}$ ($1/e^{2}$ radius) and has a power of  $\sim1~\mu\mathrm{W}$. The coupling beam has a waist size of $\sim100~\mu\mathrm{m}$ and a power of $\sim35~\mathrm{mW}$. The polarizations of the coupling and probe beams are both linear and parallel to the magnetic field along $z$. During the experiment, both probe beams in the two channels are frequency-modulated by the same acousto-optical modulator. The saturated absorption signals are demodulated and used to lock the EIT probe beam to one of the 5S$_{1/2}$ to 5P$_{3/2}$ transitions shown in Fig.~\hyperref[fig:2]{\ref*{fig:2}(a)}.

The frequency of the Rydberg-EIT coupling laser is linearly scanned
over a range of 4.5 GHz at a repetition rate of $\sim1$~Hz. The scans are linearized to within a $1~\mathrm{MHz}$ residual uncertainty using the transmission peaks of a temperature-stabilized Fabry-Perot cavity. Meanwhile, the coupler laser is chopped at $33~\mathrm{kHz}$. The EIT transmission signals are recovered by a digital lock-in referenced to the chopping frequency.

\section{RYDBERG-EIT SPECTRA IN THE HIGH-MAGNETIC-FIELD REGIME}

\begin{figure}
\includegraphics[width=\columnwidth]{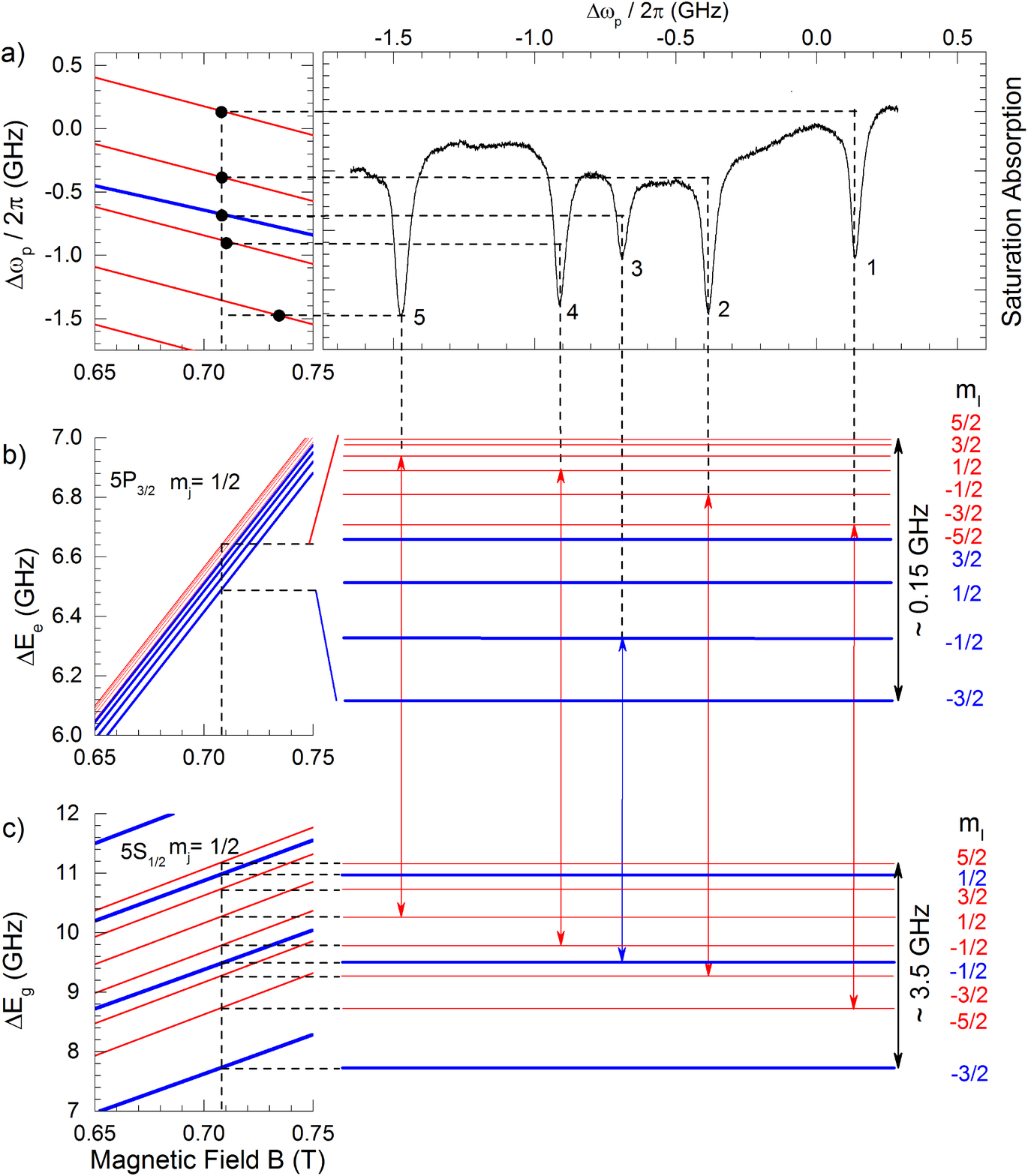}
\caption{(color online) (a) Saturation absorption signal in a $0.71$ T magnetic field versus probe frequency detuning $\Delta \omega_p$(right panel) and calculated magnetic-field-induced level shifts (left panel). The frequency is measured relative to the magnetic-field-free $^{87}\mathrm{Rb}$
$5S_{1/2},F=2$ to $5P_{3/2},F=3$ transition. (b) Schematic of atomic energy levels for intermediate state and (c) ground state. The states are labeled by the quantum numbers $m_{i}$ and $m_{j}$ (which are good quantum numbers in Pasch-Back regime). Energy levels of $^{85}\mathrm{Rb}$ and $^{87}\mathrm{Rb}$ are shown in thin red and bold blue, respectively.
\label{fig:2}}
\end{figure}

In a magnetic field of $\sim1~\mathrm{T}$,  all atomic energy levels involved in the Rydberg-EIT cascade scheme are shifted by up to several tens of GHz. The relevant ground- (5S$_{1/2}$) and intermediate-state (5P$_{3/2}$) energy levels and calculations of their field-induced shifts are plotted in Figs.~\hyperref[fig:2]{\ref*{fig:2}(b)} and~\hyperref[fig:2]{\ref*{fig:2}(c)}. In order to frequency-stabilize the EIT probe laser  to a 5S$_{1/2}$ to 5P$_{3/2}$ transition in the strong magnetic field, we implement a saturation absorption measurement using the Sat. Spec. channel, as illustrated in Fig.~\ref{fig:1}. The right panel of Fig.~\hyperref[fig:2]{\ref*{fig:2}(a)} shows the measured saturated absorption signals. Over the displayed probe frequency range, the spectrum consists of four $^{87}\mathrm{Rb}$ lines (peaks 1, 2, 4, and 5) and a $^{85}\mathrm{Rb}$ line (peak 3). Note that in the Paschen-Back regime cross-over dips are not present, because the $m_{I}$ quantum number is conserved in all optical transitions, and the separations between fine structure transitions with different $m_{j}$ exceed the Doppler width.

Due to the differences in the hyperfine coupling of $^{87}\mathrm{Rb}$ and $^{85}\mathrm{Rb}$ (e.g. magnetic dipole coupling strength, electric quadrupole coupling strength, nuclear spins and isotope shifts), the energy-levels of each isotope exhibit different shifts in the strong field. This leads to magnetic-field-dependent frequency separations $\delta_{23}$ between peaks 2 and 3, and $\delta_{34}$ between peaks 3 and 4 in the spectrum.  The ratio $\delta_{34}/\delta_{23}$ is fairly sensitive to the magnetic field. Using this feature, which relies on the presence of both isotopes in the cell, we determined the magnetic field strength sampled along the Sat. Spec. channel to be $0.71~\mathrm{T}$. This is indicated by the vertical dashed line in the left panel of Fig.~\hyperref[fig:2]{\ref*{fig:2}(a)}. At the end of the scan range, the peak positions deviate slightly from their locations expected for $0.71~\mathrm{T}$. This is caused by a slight nonlinearity of the mechanical-grating scan of the external-cavity diode laser at the end of its scan range. This nonlinearity does not affect the Rydberg-EIT experiment,  discussed in the next paragraph, because the probe laser is locked to one of the transitions in Fig.~\hyperref[fig:2]{\ref*{fig:2}(a)} and has a fixed frequency.

\begin{figure*}
\includegraphics[width=\textwidth]{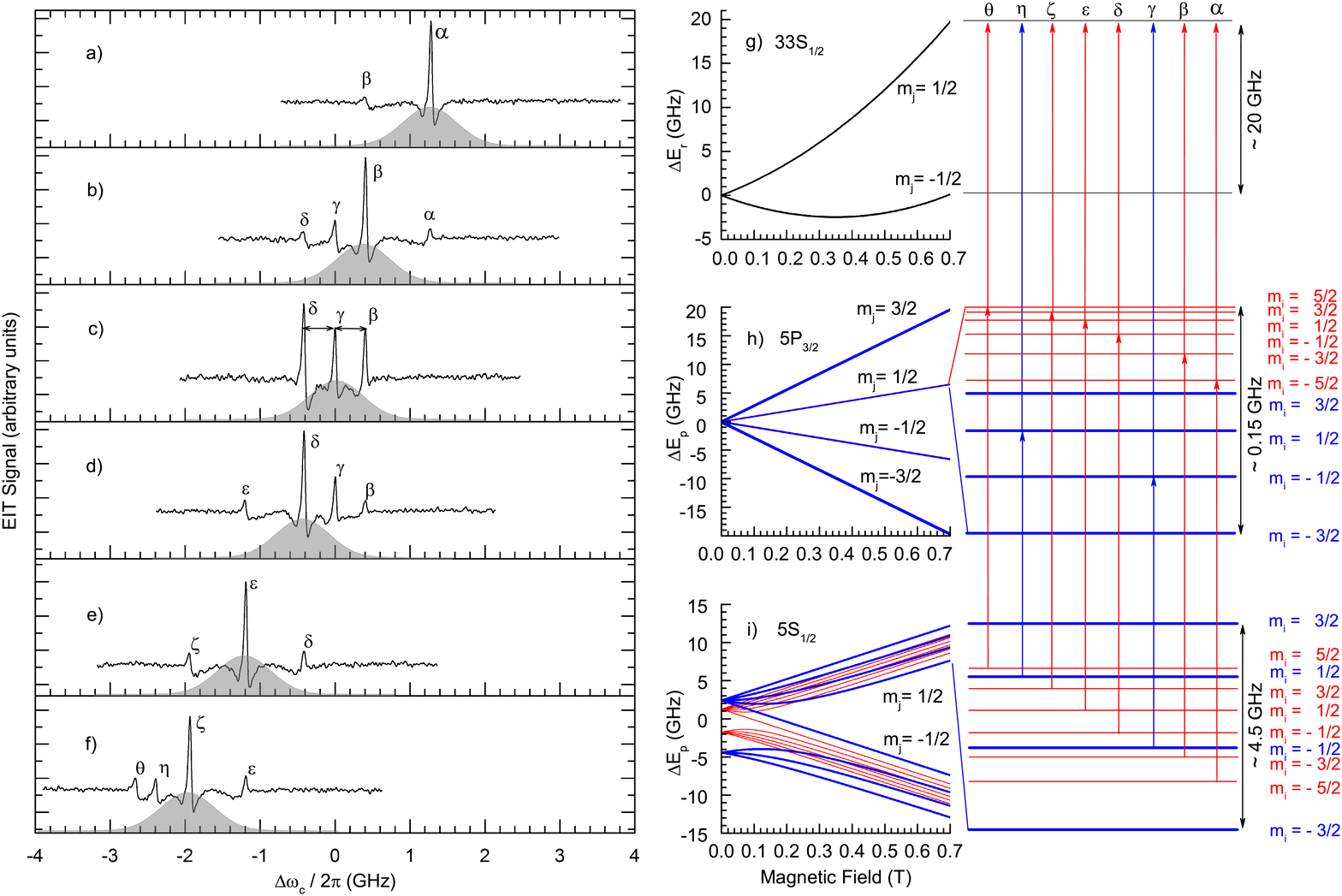}
\caption{(color online) (a) to (f) EIT transmission signals for the EIT probe
laser locked to peaks 1 to 5 in Fig.~\hyperref[fig:2]{\ref*{fig:2}(a)}.
The shaded areas indicate the weighting due to the Maxwell velocity distribution (see discussion of Eq.~(3)). Signals corresponding to the same EIT transition are labeled with the same Greek letters in each spectrum. The coupling laser detuning is given relative to peak $\gamma$ in panel (c). The consecutive spectra are shifted such that shared EIT peaks are aligned, as determined using cross-correlation functions of neighboring scans. Calculated energy level shifts in the Paschen-Back regime for the (g) Rydberg, (h) intermediate and (g) ground states. The transitions corresponding to the EIT resonances in panels (a) through (f) are indicated by vertical arrows and labeled with the same Greek letter. Transitions in $^{85}\mathrm{Rb}$ and $^{87}\mathrm{Rb}$
are coded with thin red and bold blue lines, respectively.
\label{fig:3}}
\end{figure*}

To investigate Rydberg-EIT in the high-magnetic-field regime, we frequency-stabilize the $\pi$-polarized EIT probe laser to a $\vert 5S_{1/2}, m_j=1/2, m_i \rangle \rightarrow \vert 5P_{3/2}, m_j=1/2, m_i \rangle$ transition and access the $\vert 33S_{1/2}, m_j=1/2, m_i \rangle$ Rydberg state with a coupling laser of the same polarization. Figures~\hyperref[fig:3]{\ref*{fig:3}(a-f)} show the Rydberg-EIT spectra measured in a $B$ field of $0.70~\mathrm{T}$. Each spectrum corresponds to a different probe frequency, with the probe laser locked to a different saturation absorption peak in Fig.~\hyperref[fig:2]{\ref*{fig:2}(a)}. In each spectrum, the EIT resonances are labeled by Greek letters corresponding to different $m_{i}$ states of $^{85}\mathrm{Rb}$ and $^{87}\mathrm{Rb}$ in the cascade structure, indicated in Fig.~\hyperref[fig:3]{\ref*{fig:3}(g-i)}.

For Rydberg-EIT measurements in a vapor cell, the width of the Maxwell velocity distribution of the atoms needs to be considered, as well as the Doppler effect induced by the wavelength mismatches of the probe and coupler lasers~\cite{PhysRevLett.98.113003}. It can be shown that, if an external field shifts the ground, intermediate and Rydberg levels by $\Delta E_g$, $\Delta E_e$, and $\Delta E_r$, respectively, the coupling laser detunings, $\Delta\omega_c$, at which the EIT resonances occur are given by a linear combination of all three level shifts:

\begin{equation}
\hbar\Delta\omega_{c} = \Delta E_r+\left(\frac{\lambda_{p}}{\lambda_{c}}-1\right)\Delta E_e-\frac{\lambda_{p}}{\lambda_{c}}(\Delta E_g+\hbar\Delta\omega_p)
\label{eq:1}
\end{equation}
where $\lambda_{p}$ and $\lambda_{c}$ are the wavelengths of the probe and coupling lasers, and $\Delta\omega_p$ is the probe-laser detuning of lower transition. The wavelength-dependent prefactors are deduced by requiring resonance conditions on both the lower and the upper transitions in the three-level cascade structure.

The energy-level shifts $\Delta E_g$, $\Delta E_e$ and $\Delta E_r$ in Eq.~\ref{eq:1} are plotted as a function of magnetic field in Figs.~\hyperref[fig:3]{\ref*{fig:3}(g-i)}. For S Rydberg states in Rb, which are non-degenerate and fine-structure-free, the Rydberg level shift (in atomic units) is given by~\cite{gallagher:2005rydberg}

\begin{equation}
\Delta E_r=\frac{m_sB}{2}+\frac{B^{2}}{8}\langle nlm_l\vert \hat{r}^{2}\sin^{2}\hat{\theta}\vert nlm_l\rangle \quad,
\label{eq:2}
\end{equation}
where $n, l, m_l, \mathrm{and}~m_s$ are principal, angular momentum, magnetic orbital and spin quantum numbers, respectively. The coordinates $r$ and $\theta$ are spherical coordinates of the Rydberg electron (the magnetic field points along $z$). The first term on the right hand side of Eq.~\ref{eq:2} represents the paramagnetic term of the electron spin, and the second term is the diamagnetic shift of the Rydberg state. For $\vert 33S_{1/2},m_j={1/2}\rangle$ atoms in a $1$~T field, the differential dipole moment is $3.06\mu_B$, implying that the diamagnetic contribution is about twice as large as the spin dipole moment. This fact, as well as the $\lambda_{p}/\lambda_{c}$ enhancement factor of the ground state shift ($\Delta E_g$), make the Rydberg-EIT resonances highly sensitive to small  variations in a high-magnetic-field background (see next section for details).

Eight out of the ten EIT resonances that exist for the given polarization case are present in the frequency range covered by the coupling laser in Figs.~\hyperref[fig:3]{\ref*{fig:3}(a-f)}. Every resonance satisfies Eq.~\ref{eq:1}  and has a well-defined atomic velocity, $v$, given by
  
\begin{equation}
v= \frac{\lambda_p}{2\pi}\left(\Delta\omega_p+\frac{\Delta E_g-\Delta E_e}{\hbar}\right)
\label{eq:3}
\end{equation} 
For an EIT resonance to be visible in a spectrum with given $\Delta\omega_p$, $\Delta E_g$ and $\Delta E_e$, the velocity $v$ that  follows from Eq.~\ref{eq:3} must be within the Maxwell velocity distribution. Since the probe laser is locked to one of the resonances shown in Fig.~\ref{fig:2} in every EIT spectrum, each spectrum has a strong resonance at its center for which Eq.~\ref{eq:3} yields $v\approx0$ (where the Maxwell velocity distribution peaks). For the neighboring EIT resonances, the velocities are several hundreds of meters per second, due to their large $\Delta \omega_p$. (It is seen in Fig.~\ref{fig:2} that the spacings between neighboring probe-laser resonances are several hundred MHz.) Since the rms velocity of the Maxwell velocity distribution in one dimension is about 170~m/s, the number of atoms contributing to the neighboring EIT resonances is greatly reduced relative to that of the center resonance. The finite width of the velocity distribution therefore limits the number of resonances observed in each scan to 2-4.

\section{DISCUSSION}

According to Eq.~\ref{eq:1}, the paramagnetism of the ground- and intermediate- and Rydberg-state level shifts (which are in the Paschen-Back regime) and diamagnetism of the Rydberg atoms in strong magnetic fields, lead to highly magnetic-field-sensitive shifts of the Rydberg-EIT resonances. For example,  the cascade $\vert5S_{1/2},m_{j}=-1/2\rangle\rightarrow\vert5P_{3/2},m_{j}=-1/2\rangle\rightarrow\vert33S_{1/2},m_{j}=1/2\rangle$ generates an EIT peak that shifts at 7~MHz/Gauss. The EIT resonances accessed in this work shift at about 2.5~MHz/Gauss. 

\begin{figure}=
\includegraphics[width=\columnwidth]{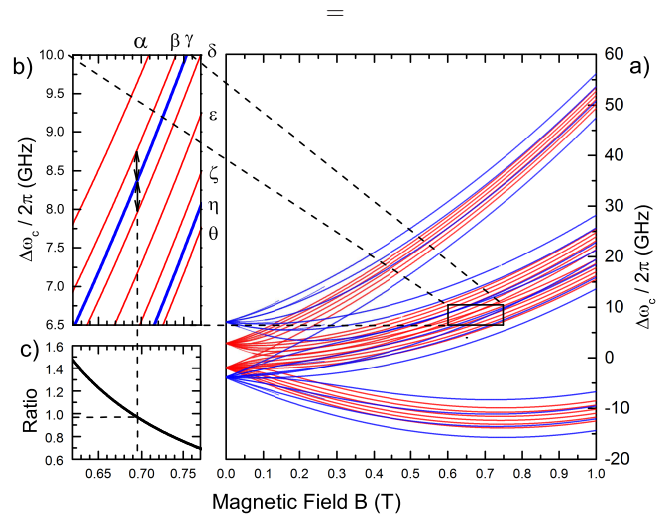}
\caption{(a) EIT line positions according to Eq.~\ref{eq:1} as a function of $B$ for $\pi-\pi$ or $\pi-\sigma$ transitions from $\lvert5S_{1/2}\rangle$ through $\lvert5P_{3/2}\rangle$ to $\lvert33S_{1/2}\rangle$ for $^{85}\mathrm{Rb}$ (thin red lines) and $^{87}\mathrm{Rb}$ (bold blue lines). (b) Zoom-in of the transitions ($\alpha$ to $\theta$) observed in this work. The two arrows indicate the frequency separations, the ratio of which we have used to extract the magnetic field strength. (c) Splitting ratio between $\beta-\gamma$ and $\gamma-\delta$ as a function of $B$
\label{Fig:4}}
\end{figure}

In Fig.~\ref{Fig:4}, we display a calculation of magnetic-field induced Rydberg-EIT resonance shifts of both $^{85}\mathrm{Rb}$ (thin red lines) and $^{87}\mathrm{Rb}$ (bold blue lines). In the high field regime, it is evident from Eq.~\ref{eq:1} and Fig.~\ref{fig:3} that the frequency separations between $^{85}\mathrm{Rb}$ and $^{87}\mathrm{Rb}$ are dependent on the magnetic field strength. This feature arises from the Paschen-Back behavior of the $\vert5S_{1/2}\rangle$ and $\vert 5P_{3/2}\rangle$ levels. The two arrows in Fig.~\hyperref[Fig:4]{\ref*{Fig:4}(b)} indicate the frequency splittings, the ratio of which we have used to extract the magnetic field strength. We map the splitting ratio from the experimental data in Fig.~\hyperref[fig:3]{\ref*{fig:3}(c)} (horizontal arrows), using the function shown in Fig.~\hyperref[Fig:4]{\ref*{Fig:4}(c)}, onto a magnetic field of $B=0.6960\pm0.0008~\mathrm{T}$. The field uncertainty arises from the spectroscopic uncertainty of the peak centers in the spectrum. From Fig.~\ref{Fig:4} it is also evident that this measurement procedure for the magnetic field relies on the presence of both Rb isotopes in the cell.

The spectra in Fig.~\ref{fig:3} and Fig.~\ref{Fig:5} below show that the EIT lines exhibit an asymmetric structure. This is in part due to the field inhomogeneities within the measurement volume. The inhomogeneities affect the line width (line-broadening), shift the line centers (line-pulling), and cause the characteristic triangular shape of the EIT resonances. The origin of these effects needs to be reasonably well understood to confirm the accuracy of the above magnetic-field measurement.

In order to quantitatively model the spectra, we use a Monte Carlo simulation to find the total power loss of the probe beam due to the photon scattering by the atoms in the inhomogeneous magnetic field. The atoms are excited by laser beams with Gaussian mode profiles. To model the atomic response, the steady-state of the excited-state population is calculated using the Lindblad equation for the three-level cascade structure~\cite{PhysRevA.85.033830} with position-dependent Rabi frequencies and magnetic-field-dependent level shifts. In the simulation, we randomly pick the atomic positions $\bf{R_i}=(X_i,Y_i, Z_i)$ from a uniform distribution truncated at the cell boundaries, and velocities in $x$ direction from a one-dimensional Maxwell velocity distribution for 300~K ($i$ is the atom counting index). The $B$ fields at positions $\bf{R_i}$ are given by the results of the FEM field calculation shown in Fig.~\ref{fig:1} and in the inset of Fig.~\hyperref[Fig:5]{\ref*{Fig:5}(c)}. The field-induced energy-level shifts are taken from data sets used in Figs.~\hyperref[fig:3]{\ref*{fig:3}(g-i)}.  The probe and coupler Rabi frequencies, $\Omega_p(\bf{R_i})$ and $\Omega_c(\bf{R_i})$, are determined by the beam parameters given in Sec.~\ref{sec:2} with center Rabi frequencies, $\Omega_{p0} =2\pi \times 11~\mathrm{MHz}$ and $\Omega_{c0}=2\pi \times 6.8~\mathrm{MHz}$. Further, we consider the natural isotopic mix and assume a uniform distribution of the atoms over all $m_i$ states. The probe detuning is set to $\Delta \omega_p =(\Delta E_e -\Delta E_g)/\hbar$ for the peak $\delta$ at $B=0.6960$T, and the coupler detuning is varied. The spectrum is simulated using a sample of $10^6$ randomly selected atoms. The averaged simulated spectrum is shown in Fig.~\hyperref[Fig:5]{\ref*{Fig:5}(b)}. 

The simulation agrees very well with the experimental spectrum, as shown in Fig.~\hyperref[Fig:5]{\ref*{Fig:5}(a,b)}. In the present experiment, the $B$ field inhomogeneity (see Fig.~\ref{fig:1}) dominates the line broadening. The magnetic-field variation in the probe volume is about 35 Gauss, which corresponds to a line broadening of $\sim 100~\mathrm{MHz}$ (vertical dashed lines in Fig.~\ref{Fig:5}). The simulation also reveals that the line centers are pulled by the same amount of $-10~\mathrm{MHz}$ relative to the theoretical line positions expected for the maximum magnetic field. Therefore, the ratio of the frequency separations indicated by the arrows in Fig.~\hyperref[Fig:5]{\ref*{Fig:5}(b)}, which we have used to determine the magnetic field in Fig.~\hyperref[Fig:4]{\ref*{Fig:4}(c)}, is unaffected by the line-pulling. 

\begin{figure}
\includegraphics[width=\columnwidth]{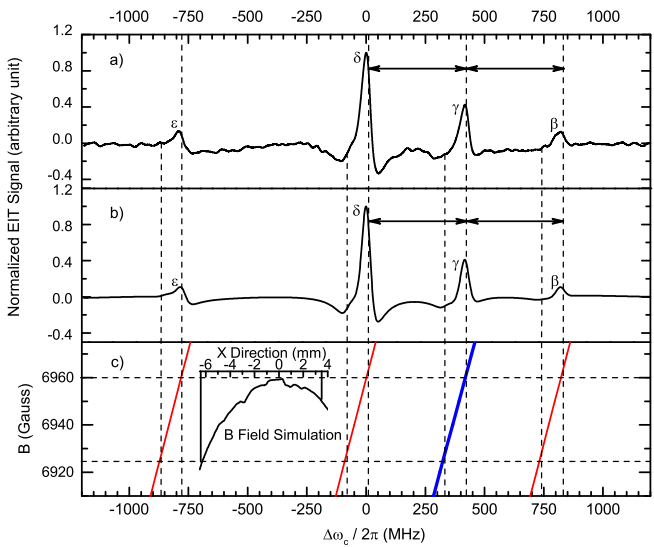}
\caption{(a) Zoom-in look of Fig.~\hyperref[fig:3]{\ref*{fig:3}(d)}. (b) Monte
Carlo simulation of the EIT spectrum shown in (a). (c) The magnetic field induced EIT resonance shifts that contribute to the spectroscopic line-broadening and pulling for peaks $\epsilon$, $\delta$, $\gamma$, and $\beta$. The vertical
dashed lines indicate the line shape extension range resulting from the inhomogeneous magnetic field inside the cell. (Power broadening is on the order of 20~MHz). The
inset shows FEM simulated magnetic fields distribution along $x$ axis within the cell boundaries indicated by vertical black lines.
This distribution is off centered by 1.5 mm to reflect possible asymmetry
in the experimental setup. \label{Fig:5}}
\end{figure}

The only free parameter in the simulation is the decoherence rate of the Rydberg state. We have found that this parameter has a profound effect on the depth of the side-dips next to all EIT peaks. In order to explain the experimentally observed spectra, we have to assume a Rydberg dephasing rate of $2\pi \times 50$~MHz, with an uncertainty of $\pm10 ~\mathrm{MHz}$. This dephasing rate is unexpectedly high, when compared to other Rydberg-EIT and Aulter-Townes work~\cite{PhysRevA.93.022709,PhysRevA.90.043849}. This large dephasing rate might be due to free charges generated by Penning and thermal ionization of Rydberg atoms and magnetic trapping of the charges~\cite{PhysRevA.86.023416}. The origin of the dephasing is currently under investigation.

In the simulation we ignore optical pumping from the intermediate $\vert5P_{3/2},m_j=1/2\rangle$ into the ground level $\vert5S_{1/2},m_j=-1/2\rangle$. We believe this is justified by the short atom-field interaction time ($\sim200~\mathrm{ns}$), which ensures only a few scattered photons per atom. We note that any optical pumping effects will only lead to a global attenuation of the EIT line strengths. Further, the EIT leads to a reduction in the probe-photon scattering rate, modifying the optical pumping near the EIT resonances~\cite{Linjie:arXiv:1702.04842}. In our case this is not expected to substantially alter the EIT line shapes. It is noted that optical-pumping effects could be, in principle, entirely avoided by selecting the $\vert 5S_{1/2},m_j=1/2,m_i\rangle$ $\rightarrow \vert 5P_{3/2},m_j=3/2,m_i\rangle$ transition for the probe laser.

An important feature of Rydberg-EIT in strong magnetic fields is the large diamagnetism of the involved Rydberg state. The diamagnetic enhancement enables the detection of small variations in a large magnetic field. Since the diamagnetic contribution to the differential dipole moment scales as $n^4 \times B$, the sensitivity of this measurement actually increases with the strength of the background field, and it can also be vastly increased by going to higher principal quantum numbers. We note that in sufficiently high $B$ fields and large enough $n$ the Rydberg-atom spectrum becomes ``chaotic''~\cite{friedrich2006theoretical}. The resultant added complexity of the spectrum will make Rydberg-EIT even more sensitive to minute variations in large $B$ fields.

\section{CONCLUSION}
In this work, we have studied vapor-cell Rydberg-EIT in a strong magnetic field, in which ground-,  intermediate- and Rydberg-states are all in the Paschen-Back regime. By exploiting the differential magnetic-field-induced shifts of the $^{85}\mathrm{Rb}$ and $^{87}\mathrm{Rb}$ EIT lines, we have measured a magnetic field of $0.6960$~T with a  $\pm 0.12\%$ uncertainty. Simulated and observed spectra show excellent agreement.  The spectra indicate an unusually large Rydberg-state dephasing rate, the origin of which we intend to explore in future work. Further, the large differential magnetic dipole moment of the diamagnetic Rydberg levels, which scales as $B$, suggests that the method holds promise for high-precision absolute and differential measurements of strong magnetic fields. By extending the work to even larger magnetic fields and higher quantum numbers, one may also explore Rydberg-atom physics in the chaotic, high-magnetic-field regime in vapor cell experiments.

\begin{acknowledgments}
The work was supported by the NSF (PHY-1506093 and IIP-1624368) and Rydberg Technologies LLC.
\end{acknowledgments}

\bibliographystyle{apsrev4-1}
\bibliography{paper}

\end{document}